\documentclass[nofootinbib,reprint,onecolumn,amsmath,amssymb,notitlepage,aps]{revtex4-1}
\usepackage{graphicx}
\usepackage[T1]{fontenc}
\usepackage{dcolumn}
\usepackage{bm}

\usepackage{hyperref}
\usepackage{epstopdf}
\hypersetup{colorlinks, linkcolor={DarkRed},citecolor={NavyBlue},urlcolor={NavyBlue}}

\usepackage[dvipsnames, svgnames, x11names]{xcolor}
\usepackage{float}

\begin{document}

\title{Weinhold geometry and thermodynamics of Bardeen AdS black holes}

\author{Yang Guo  }\email{ guoy@mail.nankai.edu.cn}
\author{Yan-Gang Miao}\email{Corresponding author: miaoyg@nankai.edu.cn}

\affiliation{ School of Physics, Nankai University, Tianjin 300071, China}


\begin{abstract}

 Thermodynamics of Bardeen AdS black holes has attracted a great deal of attentions due to its intrinsic complications and  rich phase structures. However, the entropy and thermodynamic volume are incorrect in some literatures. In this paper we revisit the thermodynamics of Bardeen AdS black holes and provide the correct entropy and thermodynamic volume. Furthermore, thermodynamic geometries are a powerful tool to probe the microstructure of black holes. Based on the Hessian matrix of black hole mass, we introduce a thermodynamic metric  and  give its scalar curvature in Weinhold's geometry. The conformal relation between Weinhold's geometry and Ruppeiner's geometry will be changed due to the specific first law of thermodynamics for regular black holes like the Bardeen AdS black hole. We also investigate the critical behaviour of phase transitions in an extended phase space, and find that the critical behaviour of  the Bardeen AdS black hole coincides with that of liquid-gas systems.  In particular, based on the Weinhold geometry we unveil a repulsive interaction in the microstructure of the Bardeen AdS black hole under its small volume state, while it is known that only attractive interaction exists in the microstructure of the van der Waals fluid.

\end{abstract}

\maketitle
\section{Introduction}
\label{sec:intro}
In the past decades, a complete statistical mechanics that describes the microstructure of black holes was still lacking although there was a wealth of work in the thermodynamics of black holes. A geometric description of thermodynamic systems may provide an alternative attempt to probe the microstructure of black holes.
Weinhold developed~\cite{weinhold1975metric} the mass (energy) representation by constructing a linear vector space with a full metric structure, where the vector space can be constructed directly from empirical laws of thermodynamics.  For each extensity $X^{\mu}$ of a thermodynamic system, its conjugate field variable $Y_{\mu}$ is defined by
\begin{eqnarray}
	Y_{\mu}\equiv\partial M /\partial X^{\mu},
\end{eqnarray}
which constitutes a thermodynamic phase space together with $X^{\mu}$, and this phase space is required to have the properties of a vector space.
Thus, an abstract space having the full properties of inner product on the thermodynamic phase space is given,  which implies that  the thermodynamic laws can show the underlying  structure of a geometric object.
Based on the fluctuation theory of thermodynamics, on the other hand, Ruppeiner introduced~\cite{1979PhRvA..20.1608R} the entropy representation, which shows that thermodynamic systems can be described by Riemannian manifolds. Quite  interesting is that there exists~\cite{1984JChPh..80..436S} a conformal equivalence between Weinhold'geometry and Ruppeiner's geometry. In information geometry, a parameterized statistical model is considered as a Riemannian manifold in which the corresponding probability distributions can be defined. In this view, Ruppeiner's geometry is one of information theories with the  probability distribution~\cite{Aman:2003ug},
\begin{eqnarray}
	\mathcal{P}(X)\propto \sqrt{g^{\rm R}}\, e^{-\frac{1}{2}g^{\rm R}_{\mu\nu}X^{\mu}X^{\nu}},
\end{eqnarray}
where $g^{\rm R}_{\mu\nu}$ is the Ruppeiner metric and ${g^{\rm R}}$ is the determinant of the metric.

Thermodynamic geometries have been extensively studied and applied~\cite{Aman:2006mn,Aman:2005xk,Shen:2005nu,Cai:1998ep,Mansoori:2014oia,HosseiniMansoori:2019jcs} to numerous systems.
In particular,  the interpretations of thermodynamic curvatures are linked~\cite{1999JPhA...32.6373O}  to the interactions in the microstructure of ideal quantum gases. In recent years, Ruppeiner's geometry has promoted substantially the connection between microstructure and thermodynamics of black holes, furthermore, such a connection has been shown~\cite{Niu:2011tb,Wei:2019yvs,Dehyadegari:2020ebz}
to be  analogous to that in the van der Waals fluid.
These interpretations are new attempts to extract information from the thermodynamic geometry of black holes.
The thermodynamic curvature does not depend~\cite{Ruppeiner:2013yca} on the thermodynamic metric since it is an invariant for a given thermodynamic state, which has recently been verified~\cite{Xu:2019gqm} again in two new phase spaces.

As is known, there is only an attractive interaction in the microstructure of van der Waals fluids. However,  there exists~\cite{Wei:2019uqg} a weak repulsive interaction in the microstructure of Reissner-Nordstr\"om AdS black holes. On the other hand, the current thermodynamic geometries are generally based on the first law of black hole mechanics associated with Maxwell's theory, so they cannot be applied  directly to interpret the microstructure of those regular black holes that associate with nonlinear electrodynamics.
We need to construct the thermodynamic geometry in the context of nonlinear electrodynamics, which is able to give an interpretation of microstructure for regular black holes. It is challenging to generalize thermodynamic geometry for understanding the regular black hole's microstructure because  regular black holes have the intrinsic complications in their corresponding first laws and Lagrangians that are coupled with nonlinear electrodynamics instead of linear one. The intrinsic complications of the Bardeen AdS black hole in the contexts of its coupling with nonlinear electrodynamics and its rich thermodynamic phase structures inspire~\cite{2019PhLB..788..219T,Singh:2020rnm,Rizwan:2020bhp,Kumara:2021hlt}  a great deal of attentions in various aspects of thermodynamics. However, thermodynamic quantities of the Bardeen AdS black hole such as entropy and thermodynamic volume in these literatures are incorrect. Although the entropy is assumed to obey the area law in Refs.~\cite{Guo:2021zxl,Pu:2019bxf}, no proper derivations are given. Consequently, we shall give the correct entropy and thermodynamic volume of Bardeen AdS black holes and then construct the Weinhold geometry for interpreting the microstructure of Bardeen AdS black holes.

The outline of this paper is as follows. In Sec.~\ref{sec:geometry} we review briefly the  properties of thermodynamic geometry and then derive the general form of Weinhold scalar curvatures. Next, in Sec.~\ref{sec:thermodynamics} we give the correct entropy and thermodynamic volume of the Bardeen AdS black hole in terms of the specific first law associated with this type of regular black holes, and demonstrate the critical behaviour of phase transitions in Bardeen AdS black holes and the analogous behaviour in  liquid-gas systems. In Sec.~\ref{sec:Waals} we provide a Weinhold geometric approach as a test to revisit the attraction in microscopic molecules of van der Waals fluids. In Sec.~\ref{sec:geometry of Bardeen} we compute the Weinhold scalar curvature of Bardeen AdS black holes and compare it with that of van der Waals fluids. In particular, we point out a repulsive interaction in the microstructure of Bardeen AdS black holes by analyzing the extreme phenomena and distributions of the Weinhold and Ruppeiner scalar curvatures. Finally, we give our conclusions in Sec.~\ref{sec:con}.	

\section{General properties of thermodynamic geometry}
\label{sec:geometry}

\subsection{Thermodynamic metrics}
We start with reviewing some general properties of thermodynamic geometry theories. The thermodynamic geometry links the statistical mechanics to thermodynamics, in which an appropriate line element is crucial in the equilibrium state space of a thermodynamic system. In the Ruppeiner theory~\cite{1979PhRvA..20.1608R,Ruppeiner:1995zz}, since the thermodynamic fluctuation theory originates from statistical mechanics, certain properties of thermodynamics and statistical mechanics for a thermodynamic system are encoded into a single thermodynamic manifold  equipped with the line element,
\begin{eqnarray}
	ds^2_{\rm R}=g^{\rm R}_{\mu\nu}dX^\mu dX^\nu,
\end{eqnarray}
where the independent variables $X^\mu$ with the Greek index, $\mu=0,1,2,\cdots$, denote the extensive quantities of a thermodynamic system and  $g^{\rm R}_{\mu\nu}$ is the so-called Ruppeiner metric defined by the Hessian matrix of thermodynamic entropy,
\begin{eqnarray}
          g^{\rm R}_{\mu\nu}=-\frac{\partial^2S(X)}{\partial X^\mu \partial X^ \nu}.
\end{eqnarray}
Here $S(X)$ is the entropy of the thermodynamic system with  Boltzmann's constant $k_{\rm B}=1$, and  $g^{\rm R}_{\mu\nu}$ must be a positive definite matrix\footnote{Thermodynamic metrics must be positive definite since the entropy has~\cite{Ruppeiner:2013yca} a maximum value in equilibrium. This is the condition that ensures a thermodynamic stability. However, the positive definiteness may fail~\cite{Xu:2019gqm} due to the non-independence between entropy and volume. It is required~\cite{Vetsov:2018dte,Dimov:2019fxp} to impose Sylvetser's criterion for a global thermodynamic stability.} in order to ensure a thermodynamic stability. Furthermore, we take note of a related metric which is defined by the Hessian matrix of black hole mass, known as the Weinhold metric~\cite{weinhold1975metric},
\begin{eqnarray}
	g^{\rm W}_{\mu\nu}=\frac{\partial^2M(X)}{\partial X^\mu \partial X^ \nu},
	\end{eqnarray}
which can be obtained from the Ruppeiner metric via a transformation of fluctuation coordinates. Technically, the two metrics that describe  Ruppeiner's and Weinhold's geometries, respectively, are conformally related~\cite{1984JChPh..80..436S} to each other,
\begin{eqnarray}
	ds^2_{\rm R}=\frac{1}{T}ds^2_{\rm W},\label{Conformal}
\end{eqnarray}
where $T$ is the temperature of the thermodynamic system under consideration.

\subsection{First law of black hole thermodynamics with nonlinear electrodynamics}

For a stationary black hole in nonlinear electrodynamics, the first law of black hole thermodynamics has been established~\cite{Rasheed:1997ns} as follows,
\begin{eqnarray}
	dM=TdS+\Phi dQ_e +\Psi dQ_m, \label{Rasheedlaw}
\end{eqnarray}
where $Q_e$ and $Q_m$ denote the electric charge and magnetic charge, and $\Phi$ and $\Psi$ the corresponding potentials, respectively.  It is easy to check that Eq.~(\ref{Rasheedlaw}) is valid for Born-Infeld black holes but invalid for Bardeen black holes. The reason is that only the electromagnetic invariant was assumed to be a variable in the Lagrangian~\cite{Rasheed:1997ns}, while some other parameters, such
as the mass and magnetic charge, were not treated as variables.

The inconsistency mentioned above was overcome~\cite{2018CQGra..35n5007Z} by dealing with the mass and magnetic charge as variables, and the general formula of the first law was then given, 
\begin{eqnarray}
	dM=TdS+\Phi dQ_e +\Psi dQ_m +\sum_i K_id\beta_i,\label{ZGlaw}
\end{eqnarray} 
where $\beta_i$ denotes the mass and magnetic charge of the black hole and $K_i$ is associated with the treatment to the mass and charge as variables in the Lagrangian of the systems of black holes with nonlinear electromagnetic fields. 
As a result, one can obtain the first law of Bardeen black holes by applying the general formula Eq.~(\ref{ZGlaw}) to the Bardeen black hole,
\begin{eqnarray}
	dM=TdS+\Psi dq+\mathcal{K} dq  + \Pi dM, \label{Gaolaw}
\end{eqnarray}
where $Q_m$ has been replaced by $q$ in the following contexts.
We note that the extra terms in Eq.~(\ref{Gaolaw}),\footnote{For Bardeen black holes, $\Pi$ takes the form of $1-r_{\rm h}^3(q^2+r_{\rm h}^2)^{-3/2}$ which is obviously a dimensionless parameter with the value less than one when the charge is fixed, and $\mathcal{K}$ is given in Ref.~\cite{2018CQGra..35n5007Z}. As $q$ does not appear in the thermodynamic phase space considered in the following discussions, we omit the formulation of $\mathcal{K}$ here. For more details on the first law and Smarr's formula of black hole thermodynamics in the context of nonlinear electrodynamics coupled to Einstein gravity, please refer to Refs.~\cite{Rasheed:1997ns,2016PhRvD..94l4027F,2018CQGra..35n5007Z,mo2006regular,Gulin:2017ycu,Bokulic:2021dtz,2005GReGr..37..643B}.} $\mathcal{K} dq$ and $\Pi dM$, come from $\sum_i K_id\beta_i$ of Eq.~(\ref{ZGlaw}), and that they appear due to the proposal suggested in Ref.~\cite{2018CQGra..35n5007Z}, which has nothing to do with the fulfillment of the Smarr formula. Incidentally, the similar application of Eq.~(\ref{Gaolaw}) to other black holes with nonlinear gauge fields, such as the Born-Infeld black hole, gives the first law as expected in Ref.~\cite{Gunasekaran:2012dq}. 

In an AdS spacetime, the cosmological constant is interpreted as the thermodynamic pressure, $P=-\Lambda/(8\pi)$. When $PdV$ term is included, the first law Eq.~(\ref{Gaolaw}) can be generalized to be
\begin{eqnarray}
		dM=\frac{T}{1-\Pi}dS+\sum_{i}\frac{Y_i}{1-\Pi}dX^i, \label{Gaolaw1}
\end{eqnarray}
where the Latin index takes $1, 2, 3\cdots$ $n$,  $X^i=(V,\cdots)$ and $Y_i=(-P,\cdots)$.

\subsection{Thermodynamic curvatures}
\label{subsec:thermo}
Given the first law of black hole thermodynamics Eq.~(\ref{Gaolaw1}), the thermodynamic line elements of Weinhold's geometry can be written as
\begin{eqnarray}
	ds^2_{\rm W}=\frac{1}{1-\Pi}dTdS+\frac{1}{1-\Pi}dY_idX^i.\label{dP2}
\end{eqnarray}
We can see that there exists an conformal factor $(1-\Pi)/T$ between the Weinhold  line element and Ruppeiner line element. The conformal factor in Eq.~(\ref{Conformal}) is $1/T$, now it changes to $(1-\Pi)/T$ in terms of the general formula of the first law of black hole thermodynamics. When considering the phase space coordinates $(T,X^i)$, we have
\begin{eqnarray}
	dS=\left( \frac{\partial S}{\partial T}\right)dT+\left( \frac{\partial S}{\partial X^i}\right)dX^i; \qquad
	dY_j=\left( \frac{\partial Y_j}{\partial T}\right)dT+\left( \frac{\partial Y_j}{\partial X^i}\right)dX^i\label{dP}.
\end{eqnarray}
	Substituting Eq.~(\ref{dP}) into Eq.~(\ref{dP2}) and using the relation,
	\begin{eqnarray}
	 \frac{\partial S}{\partial X^i}=-\frac{\partial Y_i}{\partial T},
	\end{eqnarray}
we obtain the line element of Weinhold's geometry,
\begin{eqnarray}
		ds^2_{\rm W}=\frac{1}{1-\Pi}\frac{C_{X^i}}{T}dT^2+\frac{1}{1-\Pi}\left( \frac{\partial Y_i}{\partial X^j}\right)_TdX^idX^j,  \label{We}
\end{eqnarray}
where $C_{X^i}\equiv T(\partial S/\partial T)_{X^i}$ is the heat capacity at a fixed $X^i$.

For a two-dimensional non-diagonal thermodynamic metric, the thermodynamic scalar curvature is given~\cite{Ruppeiner:2013yca} by
\begin{equation}
	 	R= -\frac{1}{\sqrt{g}} \left[ \frac{\partial}{\partial X^0}\left(\frac{g_{01}}{g_{00}\sqrt{g}}\frac{\partial g_{00}}{\partial X^1}-\frac{1}{\sqrt{g}}\frac{\partial g_{11}}{\partial X^0}\right) \right.  +  \left. \frac{\partial}{\partial X^1}\left(\frac{2}{\sqrt{g}} \frac{\partial g_{01}}{\partial X^0} -\frac{1}{\sqrt{g}}\frac{\partial g_{00}}{\partial X^1}-\frac{g_{01}}{g_{00}\sqrt{g}}\frac{\partial g_{00}}{\partial X^0}\right)\right],\label{nondscacur}
\end{equation}
where $g$ means the determinant, $g\equiv g_{00}g_{11}-g_{01}g_{10}$.
For the diagonal case, Eq.~(\ref{nondscacur}) reduces to
\begin{equation}
			R=\frac{1}{\sqrt{g}}\left[\frac{\partial}{\partial X^0} \left(\frac{1}{\sqrt{g}}\frac{\partial g_{11}}{\partial X^0}\right)+\frac{\partial}{\partial X^1}\left(\frac{1}{\sqrt{g}}\frac{\partial g_{00}}{\partial X^1}\right)\right].\label{Curva}
\end{equation}
 Considering Eq.~(\ref{We}), we can express Eq.~(\ref{Curva}) to be the general form of a scalar curvature  in the coordinates $(T, X)$ as follows,
\begin{eqnarray} 	
R&=&\frac{1-\Pi}{2C_{X}^2(\partial_{X}Y)^2}\bigg\{-C_{X}\bigg[(\partial_{X}C_{X})(\partial_{X,X}Y)
+T(\partial_{T,X}Y)^2 \bigg]+(\partial_{X}Y)\bigg[-(\partial_{X}C_{X})^2 \nonumber \\
& &-T(\partial_{T}C_{X})(\partial_{T,X}Y)+C_{X} \bigg(2(\partial_{X,	X}C_{X})+\partial_{T,X}Y+2T(\partial_{T,T,X}Y)      \bigg) \bigg] \bigg\},\label{CurWein}
\end{eqnarray}
where $\partial _{T, X}Y\equiv \partial^2 Y/(\partial T \partial X)$ as known in General Relativity.  We note that Eq.~(\ref{CurWein}) shares the same divergent points at $C_{X}=0$ or $\partial_{X}Y=0$ with the scalar curvature of Ruppeiner's geometry. The reason is that their corresponding metrics are conformally related to each other.
Since the thermodynamic scalar curvature $R$ diverges at $C_{X}=0$, $C_{X}$ can be regarded as a constant with the limit to zero, namely, $C_{X}\rightarrow0^+$. Such a normalization just removes the divergence of $R$ at $C_{X}=0$, but not the divergence at $\partial_{X}Y=0$, which will be shown\footnote{For the Bardeen AdS black hole, for instance, $\partial_{X}Y=0$ is just $\partial_{V}P=0$ at a fixed temperature $T$, see Eq.~(\ref{Critical}), which corresponds to the critical point of a phase transition.} in Fig.~\ref{CW} and Fig.~\ref{CR}.
Following Ref.~\cite{Wei:2019yvs} for the treatment to the divergence of Ruppeiner scalar curvatures, we redefine the normalized scalar curvature as follows:
\begin{eqnarray}
	\mathcal{R}\equiv R\,C_{X},\label{scaleRnorm}
\end{eqnarray}
and then obtain the Weinhold scalar curvature from Eq.~(\ref{CurWein}),
\begin{eqnarray}	
\mathcal{R}=\frac{(1-\Pi)[(\partial_{X}Y)\partial_{T,X}Y-T(\partial_{T,X}Y)^2
+2T(\partial_{X}Y)(\partial_{T,T,X}Y)]}{2(\partial_{X}Y)^2}\label{WCur}.
\end{eqnarray}
This is the normalized Weinhold scalar curvature that can be applied to thermodynamic systems with zero heat capacity, such as the Bardeen AdS black hole we consider here, because its divergent behaviour\footnote{There exists~\cite{Ruppeiner:1995zz,Wei:2019yvs}  a relation between the divergence of a correlation length and  the divergence of a thermodynamic curvature occurring at the critical point of a phase transition. We shall discuss the divergence of Weinhold and Ruppeiner thermodynamic curvatures, which originates from vanishing of the denominate in Eq.~(\ref{WCur}),  for Bardeen AdS black holes in detail in section~\ref{subsec:Critical}.} at zero heat capacity
 has been removed.

\section{Thermodynamics and phase transition of Bardeen AdS black holes}
\label{sec:thermodynamics}

In this section, we revisit the thermodynamics of Bardeen AdS black holes, give the correct thermodynamic quantities and investigate the phase transition via $P-V$ criticality in an extended phase space. The  action  of Bardeen AdS black holes with the cosmological constant $\Lambda$ in the four-dimensional spacetime reads~\cite{2000PhLB..493..149A}
\begin{eqnarray}
	\mathcal{S}=\frac{1}{16\pi}\int d^4x\sqrt{-\bf g}\left( {\bf R}-2\Lambda-4\mathcal{L}(F) \right),
\end{eqnarray}
where $\bf g$ is the determinant of the metric tensor in the four-dimensional spacetime, $\bf R$ is the Ricci scalar, $\Lambda$ is related to the AdS radius $l$ via the relation $\Lambda=-3/l^2$, and the Lagrangian of nonlinear electrodynamics $\mathcal{L}(F)$ is given by
\begin{eqnarray}
	\mathcal{L}=\frac{3M}{|q|^3}\left( \frac{\sqrt{2q^2F}}{1+\sqrt{2q^2F}}\right) ^{5/2},
\end{eqnarray}
where $F$ is electromagnetic invariant,  $M$ and $q$ denote mass and  charge of Bardeen AdS black holes, respectively.
In four dimensions, the line element of spherically symmetric Bardeen AdS black holes takes the form,
\begin{eqnarray}
ds^2=-f(r)dt^2+\frac{dr^2}{f(r)}+r^2d\Omega^2,
	\end{eqnarray}
with the shape function,
\begin{eqnarray}
f(r)=1-\frac{2Mr^2}{(q^2+r^2)^{3/2}}+\frac{r^2}{l^2},
\end{eqnarray}
where $d\Omega^2$ is the line element of unit two spheres. The Hawking temperature of Bardeen AdS black holes can be calculated,
\begin{eqnarray}
	T=\frac{-2q^2+r_{\rm h}^2+3r_{\rm h}^4/l^2}{4\pi r_{\rm h}(q^2+r_{\rm h}^2)},\label{Temp}
\end{eqnarray}
where $r_{\rm h}$ stands for the horizon radius which is the solution of the algebraic equation, $f(r_{\rm h})=0$.
Note that we have to use the specific first law, Eq.~(\ref{Gaolaw}), to compute the entropy of the Bardeen AdS black hole,
\begin{eqnarray}
	S=\int\frac{1-\Pi}{T}\,dM=\pi r_{\rm h}^2,\label{Entropy}
\end{eqnarray}
where the form of $\Pi$ has been given in footnote 2.
We see that the Bardeen AdS black hole obeys the area law. Moreover, we can derive the thermodynamic volume conjugate to the thermodynamic pressure,
 \begin{eqnarray}
 	V=(1-\Pi)\left(\frac{\partial M}{\partial P} \right)_{S,q}= \frac{4}{3}\pi r_{\rm h}^3.
 \end{eqnarray}
Here we notice that the entropy and thermodynamic volume have the standard  forms but do not have the additional terms given in Refs.~\cite{2019PhLB..788..219T,Rizwan:2020bhp}. The reason is that the area law follows from the properties of a Killing vector field in a black hole exterior, and since a black hole interior does not matter for the Killing structure, a regular black hole does not differ from a singular black hole in this aspect.

\begin{figure}[htbp]
	\centering
	\includegraphics[width=0.5\linewidth]{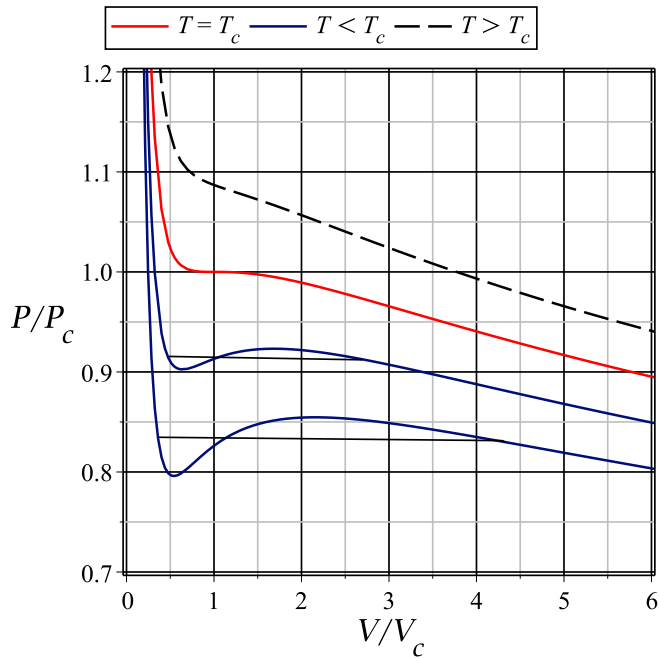}
	\caption{{\bf $P-V$ diagram of Bardeen AdS black holes.}  Isotherms descend  from top to bottom when the temperature is decreasing, where the black dashed curve corresponds to  $T>T_{\rm c}$, the red curve to the critical temperature $T=T_{\rm c}$,  and the two blue curves to $T<T_{\rm c}$.}
\label{PV}
\end{figure}

In the above considerations, the black hole mass has been identified with the enthalpy throughout the thermodynamic process in the extended phase space that includes $(P, V)$. With the clear representations of the pressure and volume, the heat capacity measuring the thermodynamic stability can be obtained,
 \begin{eqnarray}
	C_{_V}&=&T\left(\frac{\partial S}{\partial T} \right)_V=0,\\
	C_{_P}&=&T\left(\frac{\partial S}{\partial T} \right)_P=\frac{2\pi r_{\rm h}^2(q^2+r_{\rm h}^2)(-2q^2+r_{\rm h}^2+8P\pi r_{\rm h}^4)}{2q^4+7q^2r_{\rm h}^2+(-1+24P\pi q^2)r_{\rm h}^4+8P\pi r_{\rm h}^6}.
\end{eqnarray}
Unlike  the van der Waals fluid, the heat capacity of Bardeen AdS black holes at a fixed volume is zero, and the heat capacity at a fixed pressure is finite. For Bardeen AdS black holes, the equation of state $P=(T,V)$ can be derived from the Hawking temperature Eq.~(\ref{Temp}),
\begin{eqnarray}
	P=\frac{\left({\pi }/{6}\right)^{1/3} \left(8 q^2+12 T V-\left({6}/{\pi }\right)^{2/3} V^{2/3}\right)+8 \pi  q^2 T {V}^{1/3}}{12 V^{4/3}}.\label{badseqsta}
\end{eqnarray}
The behaviour of isotherms in the $P-V$ diagram is shown in Fig.~\ref{PV}.

Obviously, there exist oscillating parts (unstable regions) in the isotherms for the case of $T<T_{\rm c}$, where $T_{\rm c}$ is critical temperature, which  is similar to that of the van der Waals fluid~\cite{Kubiznak:2012wp}. For the Bardeen AdS black hole,  the horizontal segments of the isotherms\footnote{These horizontal segments are determined by Maxwell's equal area law, see ~\cite{Wei:2014qwa,Fan:2016rih} for the details. They represent the co-existence state of small and large black hole phases.} shorten when the temperature is increasing, which implies that the small and large black hole phases go to merge. This phenomenon is similar to  that happened in the van der Waals fluid where the specific volumes of gas and liquid phases approach to each other with the increasing of temperature. Moreover, the left and right ends of the horizontal segments coincide with each other when the temperature and pressure reach their critical values,  $T=T_{\rm c}$ and $P=P_{\rm c}$. This critical point is located at the inflection point of the isotherm of $T=T_{\rm c}$, which is determined by
 \begin{eqnarray}
\left( \frac{\partial P}{\partial V}\right)_T=0, \qquad \left( \frac{\partial^2 P}{\partial V^2}\right)_T=0.\label{Critical}
\end{eqnarray}
We then obtain the critical thermodynamic quantities as follows,
\begin{eqnarray}
	T_{{\rm c}}&=&\frac{\sqrt{273}+7}{\sqrt{2}    \left(\sqrt{273}+21\right) \left(\sqrt{273}+15\right)^{1/2}\pi q},\\
	V_{{\rm c}}&=&\frac{\sqrt{2}}{3}   \left(\sqrt{273} +15\right)^{3/2} \pi q^3,\\
	P_{{\rm c}}&=&\frac{219-13 \sqrt{273}}{1152 \pi  q^2}.
\end{eqnarray}
The small and large black hole phases merge~\cite{Wei:2014qwa,Fan:2016rih} at the critical point. The equation of state Eq.~(\ref{badseqsta}) can be rewritten to the reduced form,
\begin{eqnarray}
	\tilde{P}=\frac{\left(\sqrt{273}+7\right)  \tilde{T} {\tilde{V}}^{1/3}-3 \left(3 \sqrt{273}+49\right)  \tilde{V}^{2/3}+ \left(11 \sqrt{273}+189\right) \tilde{T} \tilde{V}+\sqrt{273}+21}{\left(4 \sqrt{273}+70\right)  \tilde{V}^{4/3}},\label{reeqs}
\end{eqnarray}
where the dimensionless variables are defined by
\begin{eqnarray}
\tilde{P}\equiv \frac{P}{P_{{\rm c}}},\qquad \tilde{V}\equiv \frac{V}{V_{\rm c}},\qquad \tilde{T}\equiv \frac{T}{T_{\rm c}}.\label{dimlessvar}
\end{eqnarray}
We note that the charge does not appear manifestly in the reduced equation of state Eq.~(\ref{reeqs}). It is crucial to give  such an equation of state for us to compute the thermodynamic curvatures of Bardeen AdS black holes in section~\ref{sec:geometry of Bardeen}.

\section{Weinhold geometric approach for van der Waals fluids}\label{sec:Waals}
Ruppeiner's  geometry has been well studied~\cite{Ruppeiner:1995zz,Wei:2019yvs} in different thermodynamic state spaces for van der Waals fluids.
Here we  revisit the main features of van der Waals fluids in terms of Weinhold's geometry as a warm-up
before we investigate Bardeen AdS black holes in the next section.
The equation of state of  van der Waals fluids reads
 \begin{eqnarray}
	 	\left( P+\frac{a}{v^2}\right)(v-b)=k_{\rm B}T ,
 \end{eqnarray}
 where $k_{\rm B}$ is the Boltzmann constant, $v\equiv V/N$ is the specific volume, $N$ is the number of molecules,
 and the parameter $a$ measures the attraction in the molecules of size $b$. The critical point can be fixed by Eq.~(\ref{Critical}),
 \begin{eqnarray}
 	k_{\rm B}T_{\rm c}=\frac{8a}{27b}, \qquad v_{\rm c}=3b, \qquad  P_{\rm c}=	\frac{a}{27b^2}.
 \end{eqnarray}
 Using the dimensionless variables defined by Eq.~(\ref{dimlessvar}) with the replacement of $\tilde{V}\equiv {v}/{v_{\rm c}}$,
one can rewrite the equation of state in the units of $k_{\rm B}=1$ as follows:
 \begin{eqnarray}
 	8\tilde{T}=(3\tilde{V}-1)\left(\tilde{P}+\frac{3}{\tilde{V}^2}\right).
 \end{eqnarray}
By following Ref.~\cite{weinhold1975metric} and using the above equation of state, we write the line element of Weinhold geometry in {$(T,  v)$ coordinates,
\begin{eqnarray}
		ds^2_{\rm W}&=&\frac{C_v} TdT^2-\left( \frac{\partial P}{\partial v}\right)_Tdv^2 \nonumber\\
		&=&\frac{3}{2T}dT^2-\left( \frac{2 a}{v^3}-\frac{T}{(b-v)^2}\right)dv^2. \label{vdwfwm}
\end{eqnarray}
Considering Eqs.~(\ref{Curva}) and (\ref{vdwfwm}) we obtain the Weinhold scalar curvature,
\begin{eqnarray}
	R=-\frac{2 a  (b-v)^2v^3}{3 \left(-2 a (b-v)^2+T v^3\right)^2},\label{CurvatureVanab}
\end{eqnarray}
which can also be represented in the reduced parameter space,
\begin{eqnarray}
	{R}=-\frac{4 (1-3 \tilde{V})^2 \tilde{V}^3}{3 \left[ -1+\tilde{V} \left( 6+\tilde{V}(4 {\tilde T} \tilde{V}-9)\right) \right] ^2}.	\label{CurvatureVan}
\end{eqnarray}

We plot the scalar curvature in Fig.~\ref{CWWaals} from which we can easily see the features of van der Waals fluids. Here we emphasize the following two main features. One is that the total fluid becomes a rigid body with no interactions when the specific volume of the fluid is exactly the size of each molecule ($v=b$), which is consistent with the conclusion made  by Ruppeiner's geometry in Ref.~\cite{Wei:2019yvs}. This can be seen obviously from Eq.~(\ref{CurvatureVanab}). The other feature we can deduce directly from Eq.~(\ref{CurvatureVan})  is that the Weinhold scalar curvature of van der Waals fluids is negative,  ensuring that only an attractive interaction exists in the microscopic molecules of this kind of fluids.

\begin{figure}[htbp]
	\centering
	\includegraphics[width=0.6\linewidth]{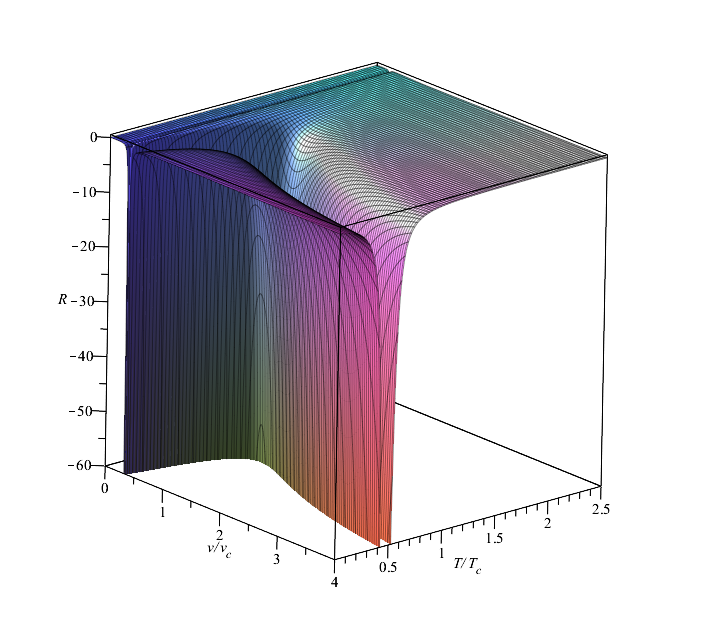}
	\caption{{\bf Weinhold thermodynamic curvature of van der Waals fluids.} This picture shows the characteristic behaviour of the Weinhold thermodynamic curvature as a function of the reduced temperature and volume.} \label{CWWaals}
\end{figure}

\section{Thermodynamic Geometry of Bardeen AdS black holes}
\label{sec:geometry of Bardeen}
Now we turn to the thermodynamic geometry of Bardeen AdS black holes which is a powerful tool to probe~\cite{HosseiniMansoori:2020jrx,Ghosh:2019pwy}  the microstructure of black holes in an extended phase space.
We shall present the results in terms of both Weinhold's and Ruppeiner's geometries that are constructed in the thermodynamic state space with the temperature and volume fluctuation coordinates, $(T, V)$.

\subsection{Weinhold's geometry}

In $(T,V)$ fluctuation coordinates, the line element Eq.~(\ref{We}) of Weinhold's geometry reads
\begin{eqnarray}
	ds^2_{\rm W}=\frac{1}{1-\Pi}\frac{C_{_V}}{T}dT^2-\frac{1}{1-\Pi}\left( \frac{\partial P}{\partial V}\right)_TdV^2,\label{barbhwg}
\end{eqnarray}
where $C_{_V}=0$ still leads to a degenerate metric which can be cured by the normalization given in Eqs.~(\ref{scaleRnorm}) and (\ref{WCur}). We note that $\Pi$ satisfies  $0<\Pi<1$, thus it alters only  the quantity rather than the sign of the thermodynamic curvature derived from the above line element. This means that the presence of $\Pi$ does not change a repulsive interaction into an attractive interaction, and vice versa. Without loss of generality, we set $\Pi=1/2$. Using Eqs.~(\ref{WCur}), (\ref{reeqs}),  and (\ref{barbhwg}), we obtain the normalized Weinhold thermodynamic scalar curvature straightforwardly,
\begin{eqnarray}
	\mathcal{R}_{\rm W}=	-\frac{21 \left[\left(79 \sqrt{273}+1305\right) \tilde{V}^{5/3}-4 \left(\sqrt{273}+15\right)\tilde{V}^{1/3}-\left(5 \sqrt{273}+83\right) \tilde{V}\right]}{\left[3 \left(\sqrt{273}+7\right) \tilde{T} \tilde{V}^{1/3}+\left(11 \sqrt{273}+189\right) \tilde{T} \tilde{V}-6 \left(3 \sqrt{273}+49\right) \tilde{V}^{2/3}+4 \left(\sqrt{273}+21\right)\right]^2},\label{CurvatureWe}
\end{eqnarray}
which is obviously  independent of the charge of black holes when the reduced thermodynamic coordinates are adopted. However, we note that it is a lengthy and cumbersome formula. This is due in large part to the intrinsic complications of the Bardeen AdS black hole in the contexts of its coupling with nonlinear electrodynamics and its rich thermodynamic phase structures~\cite{2016PhRvD..94l4027F}. We plot the behaviour of the normalized Weinhold thermodynamic curvature in Fig.~\ref{CW}.

\begin{figure}[htbp]
	\centering
	\includegraphics[width=0.75\linewidth]{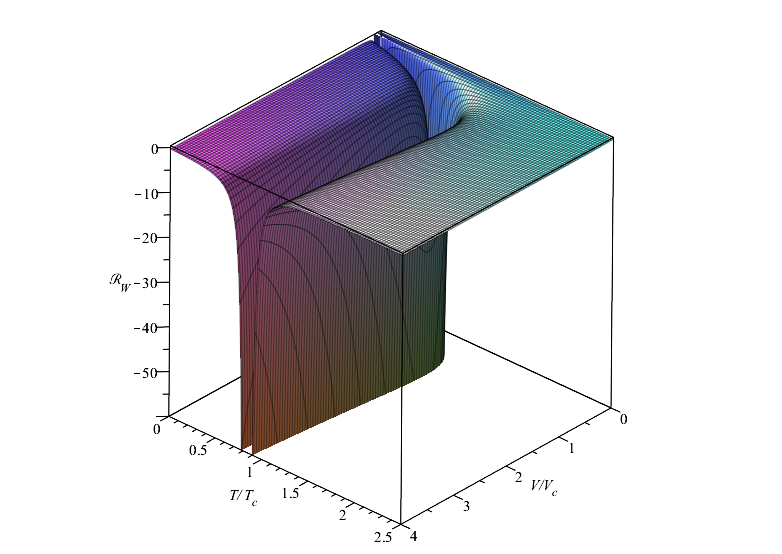}
	\caption{{\bf Weinhold  thermodynamic curvature of Bardeen AdS black holes.} This picture shows the characteristic behaviour of the Weinhold thermodynamic curvature as a function of the reduced temperature and volume.} \label{CW}
\end{figure}

The Weinhold thermodynamic curvature diverges to negative infinity when $T\rightarrow T_{\rm c}$. According to the microscopic interpretation of interactions~\cite{1999JPhA...32.6373O}, the positive and negative curvatures imply the repulsive and attractive interactions, respectively.  It is evident to see from Weinhold's curvature that there exists a strong attractive interaction in the microstructure of Bardeen AdS black holes in the vicinity of the critical temperature of phase transitions.

\subsection{Ruppeiner's geometry}
We consider the line element of Ruppeiner's geometry in fluctuation coordinates $(T, V)$~\cite{Wei:2019yvs},
\begin{eqnarray}
	ds^2_{\rm R}=\frac{C_{_V}}{T^2}dT^2-\frac{1}{T}\left( \frac{\partial P}{\partial V}\right)_TdV^2,
\end{eqnarray}
where the heat capacity vanishes at a fixed volume, i.e., $C_{_V}=T(\partial S/\partial T)_V=0$, which leads to a degenerate metric. Nonetheless, we can deal with this issue  by introducing~\cite{Wei:2019yvs} the normalized thermodynamic curvature. Following the method proposed in Ref.~\cite{Wei:2019yvs} and using the equation of state Eq.~(\ref{reeqs}), we can obtain the normalized Ruppeiner thermodynamic scalar curvature,
\begin{eqnarray}
\mathcal{R}_{\rm R}&=&\bigg[3 \left(\sqrt{273}+7\right) \tilde{T} \tilde{V}^{1/3}+\left(11 \sqrt{273}+189\right) \tilde{T} \tilde{V}-6 \left(3 \sqrt{273}+49\right) \tilde{V}^{2/3}+4 \left(\sqrt{273}+21\right)\bigg]^{-2} \nonumber \\
& &\times 84\bigg[-4 \left(\sqrt{273}+15\right) \tilde{T} \tilde{V}^{1/3}+16 \left(2 \sqrt{273}+33\right) {\tilde{V}}^{2/3}-4 \left(\sqrt{273}+17\right)\left(79 \sqrt{273}+1305\right) \tilde{T} \tilde{V}^{5/3} \nonumber \\
& &\;\;\;\;\;\;\;\;-\left(5 \sqrt{273}+83\right) \tilde{T} \tilde{V}-3 \left(21 \sqrt{273}+347\right) \tilde{V}^{4/3}\bigg].
\label{CurvatureRu}
\end{eqnarray}
We also plot the behaviour of the normalized Ruppeiner thermodynamic curvature in Fig.~\ref{CR}.

\begin{figure}[htbp]
	\centering
	\includegraphics[width=0.6\linewidth]{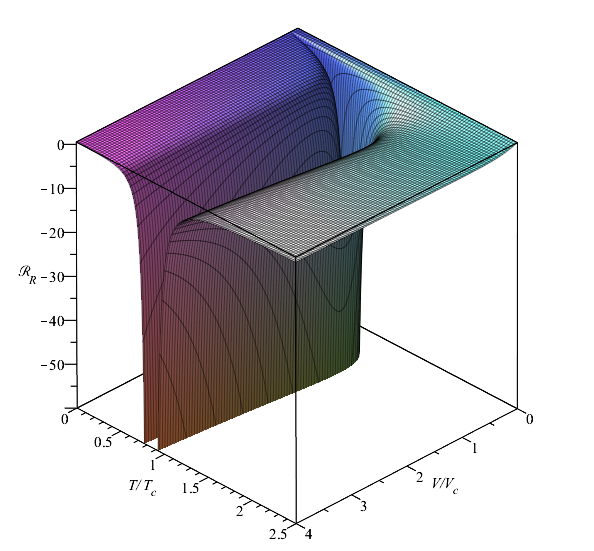}
	\caption{{\bf Ruppeiner  thermodynamic curvature of Bardeen AdS black holes.} This picture shows the characteristic behaviour of the Ruppeiner thermodynamic curvature as a function of the reduced temperature and volume.} \label{CR}
\end{figure}

From the two expressions in Eqs.~(\ref{CurvatureWe}) and (\ref{CurvatureRu}), it is obvious that the normalized thermodynamic scalar curvatures in the two geometry theories are different. There is, however, a qualitative similarity between the structures of Weinhold's geometry and Ruppeiner's geometry as displayed in Fig.~\ref{CW} and Fig.~\ref{CR}. Both the Weinhold scalar curvature and Ruppeiner scalar curvature go to negative infinity when $T\rightarrow T_{\rm c}$.  We note that the Weinhold scalar curvature and Ruppeiner scalar curvature share the same diverging curve on the $(T, V)$ plane because of the same denominator in Eqs.~(\ref{CurvatureWe}) and (\ref{CurvatureRu}).  Moreover, we point out that the Bardeen AdS black hole has a very similar microstructure to that of  van der Waals fluids~\cite{Wei:2019yvs} when the former is beyond its small volume area. We leave a detailed analysis of peculiarity of Bardeen AdS black holes in the aspect of microstructure to the next subsection when the Bardeen AdS black hole is close to its small volume area.

\subsection{Extreme phenomenon and microstructure}
\label{subsec:Critical}

In order to get a deep understanding for the normalized thermodynamic curvature of Bardeen AdS black holes, we focus on its extreme phenomena associated with infinite and vanishing curvatures. We then reveal the microstructure in which the Bardeen AdS black hole has markedly different extreme behaviours, such as  the distributions of Weinhold and Ruppeiner
scalar curvatures,  from those of the charged AdS black holes with spacetime singularities like the Reissner-Nordstr\"om AdS and Gauss-Bonnet AdS black holes~\cite{Dehyadegari:2020ebz,Niu:2011tb}. The normalized Weinhold curvature Eq.~(\ref{CurvatureWe}) and Ruppeiner  curvature Eq.~(\ref{CurvatureRu}) diverge along the following curve on the $(V/V_{\rm c}, T/T_{\rm c})$ plane,
\begin{eqnarray}
\tilde{T}|_{\mathcal{R}_{\rm W}, \mathcal{R}_{\rm R}\to \infty}&=&	\frac{6 \left(3 \sqrt{273}+49\right) \tilde{V}^{2/3}-4 \left(\sqrt{273}+21\right)}{3 \left(\sqrt{273}+7\right) \tilde{V}^{1/3}+\left(11 \sqrt{273}+189\right) \tilde{V}},\label{rinfwinf}
\end{eqnarray}
and vanish in the regions of $(V/V_{\rm c}, T/T_{\rm c})$ plane determined by
\begin{eqnarray}
\tilde{V}|_{\mathcal{R}_{\rm W}=0}=\frac{1}{12} \left(\sqrt{182}-5\sqrt{6}\right)\left(\sqrt{273}-15\right)^{1/2} ,\label{wvanish}
\end{eqnarray}
and
\begin{eqnarray}
\tilde{T}|_{\mathcal{R}_{\rm R}=0}=\frac{3 \left(21 \sqrt{273}+347\right) \tilde{V}^{4/3}-16 \left(2 \sqrt{273}+33\right) \tilde{V}^{2/3}+4 \left(\sqrt{273}+17\right)}{\left(79 \sqrt{273}+1305\right) \tilde{V}^{5/3}-4 \left(\sqrt{273}+15\right) \tilde{V}^{1/3}-\left(5 \sqrt{273}+83\right) \tilde{V}},\label{rvanish}
\end{eqnarray}
respectively. Based on the above equations we plot Fig.~\ref{TV}.

\begin{figure}[htbp]
	\centering
	\includegraphics[width=0.45\linewidth]{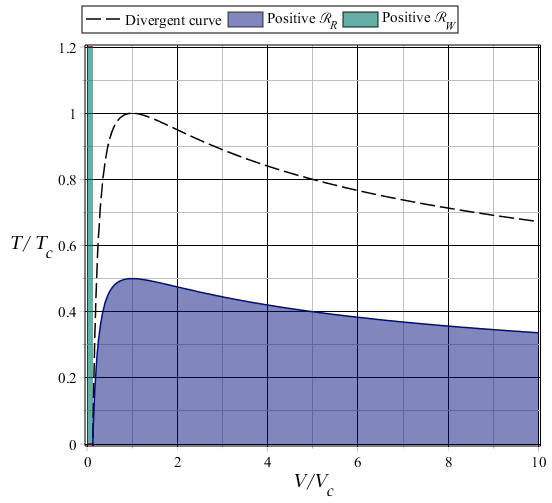}\includegraphics[width=0.45\linewidth]{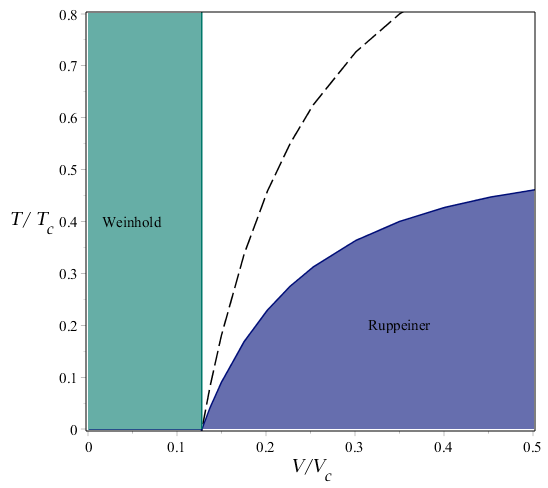}
	\caption{{\bf Extreme behaviours of thermodynamic curvatures.} (i) In the left panel the black dashed curve depicted by Eq.~(\ref{rinfwinf}) stands for the diverging case of $\mathcal{R}_{\rm W}$ and $\mathcal{R}_{\rm R}$, where the critical point of phase transitions is located at the peak. The Weinhold curvature $\mathcal{R}_{\rm W}$ is positive in the left green rectangle, where the right vertical line of the rectangle is depicted by Eq.~(\ref{wvanish}). The Ruppeiner curvature $\mathcal{R}_{\rm R}$ is positive in the bottom blue region, where the upper curve is depicted by Eq.~(\ref{rvanish}). (ii) The right panel shows the  detail features of the left panel in the small volume area.} \label{TV}
\end{figure}

We find that the Weinhold curvature shares the same diverging curve as the Ruppeiner curvature, i.e., $\tilde{T}|_{\mathcal{R}_{\rm W}\to \infty}=\tilde{T}|_{\mathcal{R}_{\rm R}\to \infty}$. This happens due to the existence of a conformal relation between the Weinhold geometry and Ruppeiner geometry. The normalized thermodynamic curvatures diverge at the critical point of a phase transition as shown by the peak of the black dashed curve in the left panel of Fig.~\ref{TV} no matter what they are described by Weinhold's geometry or Ruppeiner's geometry. We can easily see that the temperature and volume just take their critical values, $T=T_{\rm c}$ and $V=V_{\rm c}$, at this divergent point. Incidentally, a similar property of phase transition critical points was noticed~\cite{Mansoori:2013pna} for the Reissner-Nordstr\"om AdS black holes with spacetime singularities, where the critical points of a specific heat match those of a thermodynamic curvature.

In the right panel of Fig.~\ref{TV}, we zoom in  the small volume area and show the distributions of the two thermodynamic curvatures. The green and blue regions represent the positive Weinhold curvature and Ruppeiner curvature, respectively. Moreover, the Weinhold geometry and Ruppeiner geometry share the same white region in which their curvatures are negative and the only diverging curve is located in the white region. Finally, we want to emphasize that only does the Weinhold geometry not the Ruppeiner geometry of Bardeen AdS black holes present the feature of repulsive interactions in a small volume state, which does not appear in the Reissner-Nordstr\"om AdS and Gauss-Bonnet AdS black holes with spacetime singularities, and in the van der Waals fluid, either.

\section{Conclusion}
	\label{sec:con}
In the present work, we reconsider the thermodynamics and investigate the Weinhold geometry for Bardeen black holes in an anti-de Sitter spacetime. We find that the entropy of Bardeen AdS black holes still obeys the area law and the thermodynamic volume takes the standard form if the specific first law of thermodynamics, Eq.~(\ref{Gaolaw}), is considered. We also observe that the Bardeen AdS black hole has a phase transition resembling that of liquid-gas systems at its critical point, see Fig.~\ref{PV}. Moreover, we derive the normalized thermodynamic curvature in Weinhold's geometry and find that it has the same diverging curve as the Ruppeiner curvature. We also notice that  the additional parameter $\Pi$ appeared in the conformal factor of Weinhold's geometry does not reverse the interaction in the microstructure of Bardeen AdS black holes. 
In particular, we point out that the Bardeen AdS black hole as a regular black hole has a repulsive interaction in the microstructure when it is in a state of the small volume area of Fig.~\ref{TV}, which is a rather different behaviour from that of the Reissner-Nordstr\"om AdS and Gauss-Bonnet AdS black holes with spacetime singularities.

\section*{Acknowledgements}
The authors would like to thank R.-G Cai and T. Vetsov for their helpful comments and disscussions. This work was supported in part by the National Natural Science Foundation of China under Grant Nos. 11675081 and 12175108.

\bibliography{references}
\end{document}